\documentclass[12pt,a4paper]{article}

\usepackage[greek,american]{babel}
\usepackage{amssymb}
\usepackage{amsfonts}
\usepackage{amsmath}
\usepackage{relsize}
\usepackage{mathtools}
\usepackage{subfigure}
\usepackage[lofdepth,lotdepth]{subfig}
\usepackage{xcolor}
\usepackage{bm}
\usepackage[tableposition=top]{caption}
\usepackage{enumitem}

\usepackage[pdftex]{graphicx}

\usepackage{float}

\usepackage[
singlelinecheck=false 
]{caption}

\usepackage{amsthm}

\usepackage{mathtools}

\usepackage{algorithm}
\usepackage{algpseudocode}
\usepackage{pifont}

\usepackage{bbm}

\def\R{\mathbb R}

\def\E{\mathbb E}

\def\s{\sigma}
\def\t{\theta}

\def\la{\lambda}
\def\ga{\gamma}
\def\IID{\stackrel{\rm i.i.d.}\sim}

\begin{document}

\begin{center}
{{{\Large\sf\bf A Bayesian Nonparametric Approach to Dynamical Noise Reduction}}}\\
		
\vspace{0.5cm}
{\large\sf Konstantinos Kaloudis, Spyridon J. Hatjispyros
  \footnote{Corresponding author. Tel.:+30 22730 82326\\
	\indent E-mail address: schatz@aegean.gr}
}
		
\vspace{0.2cm}
\end{center}

\centerline{\sf Department of Mathematics, Division of Statistics and Actuarial Science, 
                University of the Aegean}
\centerline{\sf Karlovassi, Samos, GR-832 00, Greece.}

\begin{abstract}
We propose a Bayesian {\it nonparametric} approach for the noise reduction
of a given chaotic time series contaminated by dynamical noise, based on 
Markov Chain Monte Carlo methods (MCMC). The underlying unknown noise 
process (possibly) exhibits heavy tailed behavior. 
We introduce the Dynamic Noise Reduction 
Replicator (DNRR) model with which we reconstruct the unknown dynamic equations 
and in parallel we replicate the dynamics under reduced noise level dynamical 
perturbations. The dynamic noise reduction procedure is demonstrated specifically 
in the case of polynomial maps. Simulations based on synthetic time series are presented.
		
\vspace{0.1in}\noindent 
		
\end{abstract}
	
\section{Introduction}
\label{sec:Intro}
For over three decades, nonlinear dynamical systems \cite{ott2002chaos} have been in the center of 
attention of a wide variety of sciences, giving the opportunity to model multiple time varying phenomena, 
exhibiting complex and irregular characteristics. The unpredictable nature of chaotic dynamics was early 
connected to probabilistic and statistical methods of analysis 
\cite{berliner1992statistics, chatterjee1992chaos}. 
Furthermore, the ubiquitous effect of different kinds of noise in experimental or real data reinforced 
the interaction between nonlinear dynamics and statistics \cite{mees2012nonlinear}. 
In this context, noise reduction methods kept drawing the attention of the researchers 
from both a theoretical and an applied point of view.

Many different approaches have been adopted to address the issue of nonlinear noise reduction.  
Hammel et al. \cite{hammel1990noise}, used techniques originated from the proof of the shadowing 
lemma to reduce noise in observed chaotic data. Farmer and Sidorowich \cite{farmer1991optimal}, 
proposed the use of Lagrangian multipliers for the minimization of the distance between the 
observed and the denoised orbit. In order to deal with homoclinic tangencies, they used a 
combination of manifold decomposition and singular value decomposition techniques. Locally 
linear models were introduced by Schreiber and Grasssberger \cite{schreiber1991simple} for 
noise reduction, while Davies proposed initially gradient descent \cite{davies1993noise} and 
later Levenberg-Marquardt \cite{davies1994noise} methods for the minimization of the dynamic 
error. The first attempt to a Bayesian noise reduction framework \cite{robert2007bayesian} 
was due to Davies \cite{davies1998nonlinear}. Other methods include the usage of shadowing 
methods \cite{judd2008shadowing}, wavelet transformations \cite{jansen2012noise}, Sequential 
Markov Chain methods \cite{doucet2001introduction} in the case of state space models, and 
Kalman filtering techniques \cite{walker1997noise}, while important theoretical results 
about the consistency of signal extraction, under measurement noise, were presented by 
Lalley et al. \cite{lalley1999beneath, lalley2006denoising}.	

The type of noise contaminating the data is very important, because of the different effects induced by it.
Observational or measurement noise, originating from 
errors in the measurement process, is independent of the dynamics and can be thought of as 
being added after the time evolution of the trajectories under consideration. On the other 
hand, dynamical or interactive noise, is added at each step of the time evolution of the 
trajectories, drastically modifying the underlying dynamics. Extensive studies on the effect 
of dynamical noise on the underlying deterministic system include the works of Jaeger and Kantz
\cite{jaeger1997homoclinic}, and, Strumic and Macek \cite{strumik2008influence}. 
Dynamical noise can represent the error in the assumed model, thus compensating for a small 
number of degrees of freedom, for example a small amplitude high dimensional deterministic 
part not included in the model \cite{kantz2004nonlinear}.  Moreover, in the presence of 
dynamical noise, shadowing trajectories of non-hyperbolic maps is not possible. This problem 
was addressed by Kantz \cite{kantz1997effective}, introducing a noise reduction method based 
on ``parameter shadowing''. In this work, a shadowing pseudo-orbit is generated, evolving 
in some neighborhood of the original orbit, fulfilling the nearby rather than the exact dynamics. 	

This work regards a fully Bayesian nonparametric method for the reduction of the additive dynamical 
noise perturbing an observed noisy time series $(x_i)$ of length $n$.
We develop the DNRR model, whereby we introduce the $n$ strategic hidden random variables 
$(Y_i)$. Their posterior distribution describes all possible noise reduced trajectories
in the neighborhood of the original trajectory, and we show that with the appropriate 
point estimation, we can recover a noise reduced trajectory $(y_i)$ that for moderate 
noise levels is being generated by approximately the same dynamical system, generating the 
observed noisy time series, yet perturbed by a weaker error process. We also show that near
the homoclinic tangencies of the associated deterministic system, the posterior marginal 
distributions $Y_i$ become multimodal limiting the noise reduction levels.

The novelty of our approach lies on the fact that we make no parametric assumptions for the density of 
the noise component. 
Instead, we model the additive error using a highly flexible family of density functions, which are based on a Bayesian nonparametric model, namely the Geometric Stick Breaking process \cite{fuentes2010new}, extending previous works regarding reconstruction and prediction of random dynamical systems \cite{Hatjispyros_PhysA_Slice, Hatjispyros_CSDA_DPM, merkatas2017bayesian}. No matter what additive errors are involved, we are confident that our family of densities will be able to capture the right shape and hence statistical inference, for the parameters of interest will be improved and reliable. Under this formulation, the noise reduction method proposed can be applied to cases where the noise is not assumed to be normally distributed, or even in cases where we know that the noise component has a mixture density. Such cases include, among others, scenarios where the noise is the result of multiple sources affecting the time evolution of the underlying dynamics. In this case, our method will be able to estimate the true noise density and moreover identify the number of the sources as the ergodic average of the active clusters.

The paper is organized as follows. In Sec. II we mention some aspects of the problem and present the noise reduction algorithm steps. In Sec. III  we present the MCMC procedure for the estimation of the noise-reduced orbit. In Sec. IV we resort to simulation. We illustrate our method in the case of the random full quadratic and polynomial maps under non-Gaussian dynamical noise. We conclude in Sec. V giving some directions for further research.


\section{Preliminaries}
We define the random recurrence relation given by 
\begin{eqnarray}
\label{realproc}
X_i&=&T(\t,X_{i-1}, \ldots, X_{i-d}, e_i) \\ \nonumber
&=&g(\t, X_{i-1},\ldots,X_{i-d}) + e_i,\quad i\ge 1,
\end{eqnarray}
where $g:\Theta\times X^d\to X$, for some compact subset $X$ of $\R$,
$(X_i)_{i\ge -d+1}$ and $(e_i)_{i\ge 1}$ are real random variables over some 
probability space $(\Omega, {\cal F}, {\rm P})$; 	
we denote by $\t\in\Theta\subseteq\R^m$ any dependence of the deterministic map $g$ on parameters.
$g$ is nonlinear, and for simplicity, continuous in $X_{i:d}:=(X_{i-1},\ldots,X_{i-d})$. 
We assume that the random variables $e_i$ are independent to each other, and independent of the states $X_{i-r}$ for $r<i+d$. 
In addition we assume that the additive perturbations $e_i$ are identically distributed from a 
zero mean distribution with unknown density $f$ defined over the real line, so that 
$T:\Theta\times X^d\times\R\to\R$. Finally, notice that the lag-one stochastic process 
$(W_i^1,\ldots,W_i^d)$, formed out, from time-delayed values of the $(X_i)$ process, defined by
$$
W_i^k=\left\{
\begin{array}{lc}
g(\theta, W_{i-1}^1,\ldots,W_{i-1}^d)+e_i & \,\,\,     k=1\\
W_{i-1}^{k-1}                             & \,\,\,1<k\le d\,,
\end{array}
\right.
$$
is Markovian over $\R^d$.

We assume that there is no observational noise, so that we have at our disposal a time series
$x^n:=(x_1,\ldots,x_n)$ generated by the nonlinear stochastic process
defined in (\ref{realproc}).  The time series $x^n$ depends solely on the
initial distribution of $X_{1:d}$, the vector of parameters $\t$, and the
particular realization of the noise process.

Orbits contaminated with dynamical noise are $a$-pseudo-orbits of the underlying $g$-dynamics in the sense that for all $1\le i\le n$ there is positive $a$ for which $0<|x_i-g(\t,x_{i:d})|\le a$.
$g$-invariant measures $\mu_g(dx)$ are deformed and smoothed-out into $T$-quasi-invariant measures 
$\mu_T(dx)=\lim_{t\to\infty}{\rm P}\{x<X_t\le x+dx|\tau_{X'}>t\}$,
where $\tau_{X'}$ is a random time denoting the first passage time of the system to the unbounded
trapping set $X'=\R\setminus X$. Mind that, $\mu_T$ is not
a convolution of the unperturbed measure $\mu_g$ with the noise distribution, as it happens
in the case of observational noise. 

As a distance between the two time series $x^n$ and $y^n$, we will use
the average correction
$
E_0(x^n,y^n) = \sqrt{{1\over n}\sum_{i=1}^n(x_i-y_i)^2}.
$ 
We will measure the overall deviation of the noisy orbit $x^n$  
from the $g$-determinism, with the average dynamical error
$
E_{\rm dyn}(x^n;g) = \sqrt{{1\over n}\sum_{i=1}^{n}(x_i-g(\t,x_{i:d}))^2}.
$


\subsection{Dynamical noise reduction}

Dynamical noise has a severe effect on the underlying dynamics, i.e. the deterministic part of the noisy corrupted time series, especially when the system under consideration is non-hyperbolic. In the hyperbolic case, the shadowing lemma guarantees the existence of shadowing pseudo-orbits and moreover if the dynamical noise is bounded, it can be treated as measurement noise. This means that 
we can find a $g$-deterministic orbit 
$y^n$ and a noise process $(\tilde{z}_i)$ such that $x_i = y_i + \tilde{z}_i$.
The $\tilde{z}_i$ errors are describing the distribution of the distance between the two orbits, 
and the $x^n$-dynamical noise reduction problem can be treated as a $y^n$-observational noise 
reduction problem. This is not valid, though, when the underlying dynamics are non-hyperbolic.	

\begin{figure}
	\centering
	\includegraphics[width = 0.4\textwidth]{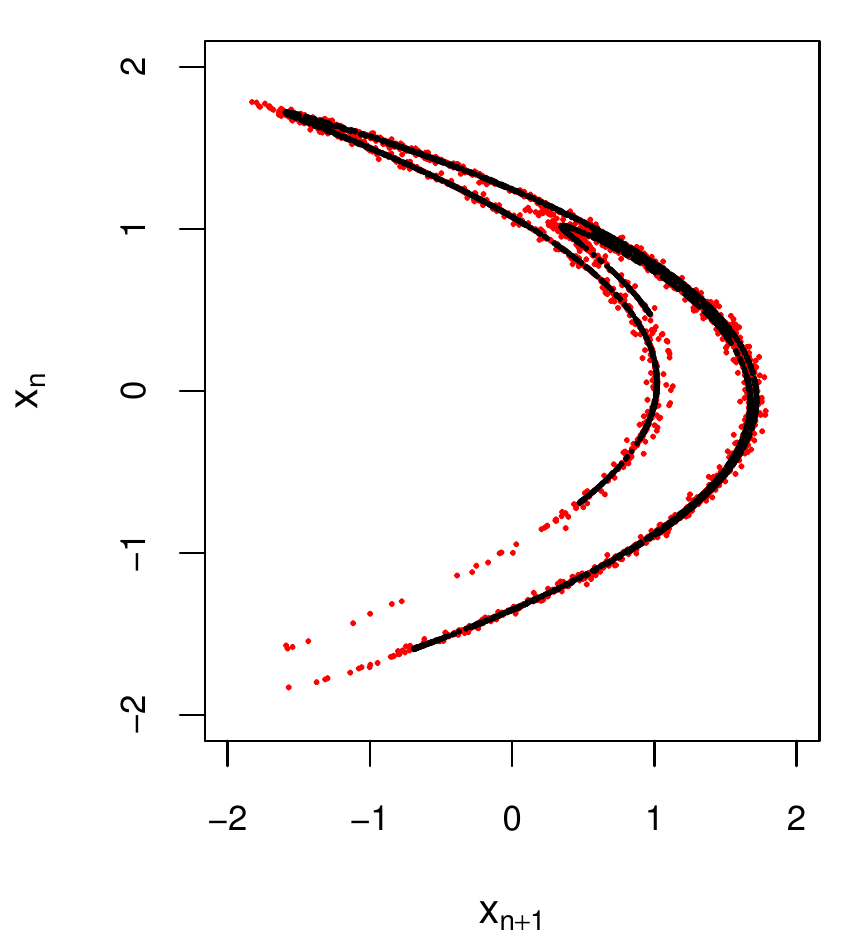}
	\caption{Noisy and deterministic H\'{e}non trajectories, of length $n=5000$, are depicted 
		in red and black, respectively, for a $3\%$ dynamical noise level.}
	\label{henons}
\end{figure}

In the non-hyperbolic case, the presence of homoclinic tangencies (HTs) in the phase space, points where the stable and unstable manifold of a hyperbolic orbit intersect tangentially, is responsible for the emergence of a much more complicated structure. In the vicinity of HT's, the dynamic perturbations are  amplifying dynamics away from the neighborhood of the attractor. One of the effects caused by the noise amplifications due to HT's are noise-induced prolongations \cite{kantz1997effective}.
For example, in Figure \ref{henons}, we display the delay plots of the deterministic and a dynamically
perturbed realization of the H\'{e}non map of lengths $n=5000$. The noisy trajectory, has been 
generated via 
\begin{equation}
\label{noisyHenon}
x_i = g(x_{i-1}, x_{i-2})=1.38 - x_{i-1}^2 + 0.27 x_{i-2} + e_i,
\end{equation}
where $e_i\IID 0.6\,{\cal N}(0,\sigma^2)+0.4\,{\cal N}(0,100\sigma^2)$, 
for $\s^2=0.21\times 10^{-4}$,
with initial condition $x_0=x_{-1}=0.5$. The time series realization has been chosen, such that, 
the noise level is approximately $3\%$. We can see the intense noise induced prolongations, as clouds of 
points in red, away from the neighborhood of the deterministic attractor (points in black).  

The deformation of the $g$-invariant measure to a
$T$-quasi-invariant measure, leads to the expansion of its support, and the perturbed map visits areas of the phase space that was not able to visit without the effect of the dynamical noise. 

We aim to reconstruct the underlying deterministic dynamics in the form of a map 
$\hat{g}_{x^n}$, and sample a $y^n$ trajectory, such that we will be able to control 
its average deviation from determinacy $E_{\rm dyn}(y^n,\hat{g}_{x^n})$,
with respect to $\hat{g}_{x^n}$, as well as its average correction $E_0(x^n,y^n)$, 
with respect to $x^n$.


\subsection{Gaussian and non-Gaussian noise processes}

We assume that the corrupting noise $f$, responsible for the observed time series $x^n$, 
can be represented as a countable mixture of zero mean normals ${\cal N}(z|0,\sigma_i^2)$ 
of variances $\sigma_i^2$, that is 
$$
f(z):=f_M(z)=\sum_{i=1}^M p_i\,{\cal N}(z|0,\sigma_i^2)
$$
with $p_i>0$ and $\sum_{i=1}^M p_i=1$, where $M$ can be infinite.
Then, the variance associated with $f_M$ (when it exists), is the $p_i$-mixture of the $\sigma_i^2$-variances i.e.
$\sigma_{f_M}^2=\sum_{i=1}^M p_i\sigma_i^2$.
Following Jaeger and Kantz \cite{kantz1997effective} we define the noise level $\eta$ as the percentage of the sampling
standard deviation of $x^n$ (the signal), that is, $\eta=100\,\sigma_f/\sigma_{x^n}$.

As a measure of the departure from normality of the noise process $f$, we use the mean absolute 
deviation from the mean, normalized by the standard deviation. So for a zero mean $Z\sim f$ it is that
$TF_f:=\E|Z|/\sqrt{\E|Z|^2}$. The closer the quantity $TF_f$ is to one, the thinner the tails are. 
We have the following lemma:

\smallskip\noindent
{\bf Lemma $1$.} {\sl For all $M\ge 1$, it is that 
	\begin{equation}
	\label{TFequation}
	TF_{f_M}\le TF_{f_1}\quad{\rm with}\quad TF_{f_M}={1\over\s_{f_M}}\sqrt{2\over\pi}\sum_{i=1}^M p_i\sigma_i.
	\end{equation}
}

\smallskip\noindent
{\bf Proof:~}
We let $|Z|\sim f_+$, then it is clear that 
$$
f_+(z)={f_M(z)\,{\cal I}(z>0)\over\int_{\R^+}f_M(z)dz}
=2\sum_{i=1}^M p_i\,{\cal N}(z|0,\sigma_i^2){\cal I}(z>0),
$$ 
where ${\cal I}(z>0)$ is the characteristic function of the interval $(0,\infty)$.
The equation for $TF_{f_M}$ in (\ref{TFequation}) can be verified by the fact that 
$\int_{\R^+}z\,{\cal N}(z|0,\s_i^2)dz=\s_i\sqrt{2/\pi}$. By Jensen's concave inequality
we have that $\sum_{i=1}^M p_i\sigma_i\le\s_{f_M}$ or equivalently that $TF_{f_M}\le TF_{f_1}$.\hfill $\square$

\smallskip
We consider the noise processes $f_1$ and $\{f_{2,l}:1\le l\le 4\}$ given by
\begin{align}
\label{fiveproc}
f_{1}(z)   = & \, {\cal N}(z|0,\s^2)\\
f_{2,l}(z) = & \, {5+l\over 10}{\cal N}(z|0,\s^2)+{5-l\over 10}{\cal N}\left(z|0,100\s^2\right).\nonumber
\end{align}
From lemma 1, irrespective of the choice of $\s^2$, it is that $TF_{f_{2,l}}<TF_{f_1}=\sqrt{2/\pi}$, and
the $TF_{f_{2,l}}$ sequence is decreasing, namely, it can be verified that 
$\{T_{f_{2,l}}:1\le l\le 4\}=\{0.58, 0.53, 0.49, 0.46\}$.

The motivation for a Bayesian {\it nonparametric} framework for noise reduction comes 
from the fact that the application of stochastic methods, under the false assumption of a 
normal noise process ($f=f_1$), will artificially enlarge the estimated variance of the presumed
normal errors, thus, causing poor inference for the system parameters of interest, as demonstrated 
in Ref. \cite{merkatas2017bayesian}.


\section{The dynamic noise reduction replicator model} 	
In this work, given a noisy corrupted time series $x^n$, we will use a Bayesian 
nonparametric approach to estimate the posterior joint density
of a noise reduced vector of random variables $Y^n=(Y_1,\ldots,Y_n)$.
A noise reduced time series $y^n=(y_1,\ldots,y_n)$, will be formed by some central
tendency statistic applied to predictive samples of the marginal posterior densities (MPDs) 
for all $i=1,\ldots,n$.
We define, the estimated vectors $\hat{\t}_{x^n}$ and $\hat{\t}_{y^n}$, of the control parameters 
of $g$, and the associated estimated noise components $\hat{f}_{x^n}$ and $\hat{f}_{y^n}$,
based on the time series $x^n$ and $y^n$, respectively.
Our intention is to create the $y^n$ time series, in such a way, that it 
possesses an underlying estimated deterministic law, ${\hat g}_{y^n}(\,\cdot\,):=g(\hat{\t}_{y^n},\,\cdot\,)$, that is in some sense  
(to be made precise in the sequel) close to the estimated deterministic law, 
$\hat{g}_{x^n}(\,\cdot\,):=g(\hat{\t}_{x^n},\,\cdot\,)$, responsible for $x^n$,
such that, the estimated noise component $\hat{f}_{y^n}$ influencing interactively the $y^n$
time series, will be a weaker version of the estimated dynamic noise component $\hat{f}_{x^n}$ 
influencing the original $x^n$ time series.
We remark that the $\hat{g}_{y^n}$ and $\hat{f}_{y^n}$ estimations under the noise reduced trajectory
$y^n$ have been produced via the GSBR-sampler in Ref. \cite{merkatas2017bayesian}. 


\subsection{A generic probability model}

To permit a stochastic approach to the estimation of the unobserved sequence $y^n$, under the generic assumption of a symmetric zero mean dynamical error process, we adopt the following stochastic model: 
\begin{align}
\label{model1}
x_i & = g(\t, x_{i:d})+e_i,\quad e_i\IID f(\,\cdot\,) \\ \nonumber
f(\,\cdot\,)& =\sum_{k=1}^\infty w_k\,{\cal N}(\,\cdot\,|0,\la_j^{-1}),\,\, 1\le i\le n \\
y_i & = g(\t, y_{i:d})+\zeta_i,\quad\zeta_i\IID{\cal N}(\,\cdot\,|0,\delta)\nonumber\\
y_{1:d} & = x_{1:d},\,\,{\rm P-a.s.}\quad{\rm and}\quad |x_i-y_i|<\ga_i,\quad\ga_i\IID h(\,\cdot\,),\nonumber
\end{align}
where we define $w^\infty=(w_k)_{k\ge 1}$ to be an infinite sequence of random probability weights, 
$\la^\infty=(\la_k)_{k\ge 1}$ an infinite sequence of independent and identically distributed (i.i.d) 
positive random variables (the precisions), with the two sequences $w^\infty$ and $\la^\infty$ independent 
of each other. The positive random variables $\ga_i$ are i.i.d. from some distribution $h$, possibly depending on parameters.

We will show numerically, that under a reasonable choice for the prior distribution of
the variable $\tau=\delta^{-1}$, the posterior distribution of $\delta$ (the variance), 
will concentrate its mass near zero.
This, will enable us, to minimize the overall deviation of the $y^n$ trajectory from
the estimated determinism.
To control the similarity of the $y^n$ trajectory, with respect to the observed $x^n$  
trajectory, we assume that both trajectories originate from the same initial point, that is,
$y_{1:d}=x_{1:d}$. At the same time, a-priori, we restrict each $y_i$ to be $\ga_i$-close to $x_i$. 
The latter statement, conveys prior 
information, on the {\it proximity} of the variable $y_i$ to the data point $x_i$.
Finally, we remark that the random mixture 
$\omega\mapsto f(\,\cdot\,,\omega)=\sum_{k=1}^\infty w_k(\omega)\,{\cal N}(\,\cdot\,|0,\la_j^{-1}(\omega))$
undertakes the r\^ole of a nonparametric prior over the noise density assumed
responsible for the time series $x^n$, supported over the space of densities with mean zero,
which are in turn supported over $\mathbb R$. 

We note the following lemma, which will prove useful in the sequel: 

\smallskip\noindent
{\bf Lemma $2$.} {\sl Letting ${\cal P}:=\cap_{i=1}^n\{|x_i-y_i|<\ga_i\}$, we have the following cases:
	\begin{enumerate}
		\item
		When $\ga_i=\bar{\ga}_i=$const. a.s. for $1\le i\le n$, it is that 
		$$
		{\rm P}\left({\cal P}|\,x^n,y^n\right)=\prod_{i=1}^n{\cal I}(x_i-\bar{\ga}_i<y_i<x_i+\bar{\ga}_i).
		$$
		\item
		If $\ga_i\IID{\cal W}(2,\sqrt{2/\rho})$ for $1\le i\le n$, where $\rho$ is a fixed hyperparameter,
		and ${\cal W}(a,b)$ denotes the Weibull distribution of shape $a$ and scale $b$, we have that
		$$
		{\rm P}\left({\cal P}|\,x^n,y^n\right)=\exp\left\{-{\rho\over 2}\sum_{i=1}^n(x_i-y_i)^2\right\}.
		$$
	\end{enumerate}
}

\smallskip\noindent
{\bf Proof:~}	(1.) When $\ga_i=\bar{\ga}_i$ for all $1\le i\le n$, it is that 
$$
{\rm P}\left({\cal P}|\,x^n,y^n\right)=
\begin{dcases}
1  & y_i\in(x_i-\bar{\ga}_i, x_i+\bar{\ga}_i),\,\,1\le i\le n\\
0  & \quad\quad{\rm otherwise}.
\end{dcases}
$$

(2.) Because $\ga_i\IID{\cal W}(2,\sqrt{2/\rho})$ if and only if $\ga_i^2\IID{\cal E}(\rho/2)$, 
where ${\cal E}(a)$ denotes the exponential distribution with mean $1/a$, it is that
$$
{\rm P}\left({\cal P}|\,x^n,y^n\right)=
\prod_{i=1}^n{\rm P}\{\ga_i^2>(x_i-y_i)^2\}=\prod_{i=1}^n\exp\left\{-{\rho}(x_i-y_i)^2/2\right\},
$$
which gives the desired result.\hfill $\square$


\subsection{The posterior model} 

We consider the posterior of the stochastic quantities 
$f$, $\t$, $x_{1:d}$, $y_{1:d}$, $\tau$ and $y^n$ given the data set $x^n$, 
the restriction ${\cal R}:=\{y_{1:d}=x_{1:d}\}$, the proximity information 
${\cal P}$, and the model space $\cal M$ for the functional representation of the 
deterministic part $g(\t,x_{i:d})$; for example, the model space
could be the ring $\R[x_{i:d}]$ of polynomial functions in the variable $x_{i:d}$,
with coefficients over $\R$. Then, using Bayes' theorem, we have
\begin{align}
\label{posterior1}
& \pi\left(f, \t, x_{1:d}, y_{1:d}, \tau, y^n|\, x^n, {\cal R},\, 
{\cal P},{\cal M}\right)\propto\\
& \quad\pi\left(f,\t, x_{1:d}, y_{1:d},\tau \right)\,
\pi\left(y^n,x^n,{\cal R},{\cal P}|
f, \t, x_{1:d}, y_{1:d}, \tau, {\cal M}\right),\nonumber
\end{align}
where $\pi(f,\t,x_{1:d},y_{1:d},\tau)$ is the prior density. Having in mind, that the 
estimation of the noise density $f$ is equivalent to the estimation of the variables 
$w^\infty$ and $\la^\infty$,
the likelihood factor on the second line of equation (\ref{posterior1}), becomes
	\begin{align}
	\label{likeli}
	& \pi\left(y^n,x^n,{\cal R},{\cal P}|w^\infty, \la^\infty, \t, x_{1:d}, y_{1:d}, \tau, {\cal M}\right) =\\
	& \quad{\rm P}({\cal R}|\, y_{1:d}, x_{1:d})\,{\rm P}\left({\cal P}|\, x^n, y^n\right)
	\pi(x^n|\, w^\infty, \la^\infty, \t, x_{1:d}, {\cal M})\,\pi(y^n|\, \t, \tau, y_{1:d}).\nonumber
	\end{align}
We believe, that it will be more efficient to control the average corrections between the two 
trajectories, under the assumption that $\ga_i^2\IID{\cal E}(\rho/2)$. For this reason, we
augment the conditional part of our posterior by the hyperparameter $\rho$. Then, 
taking into account the model representation for the noise components in 
(\ref{model1}), lemma 2, the fact that P$({\cal R}|\,y_{1:d},x_{1:d})
={\cal I}(y_{1:d}=x_{1:d})$ and the likelihood representation in (\ref{likeli}), the posterior becomes 
	\begin{align}
	& \pi\left(w^\infty,\la^\infty, \t, x_{1:d}, \tau, y^n|\, x^n, {\cal R},\, 
	{\cal P},{\cal M}, \rho\right)\propto\pi\left(w^\infty,\la^\infty,\t, x_{1:d},\tau \right)\nonumber\\
	& \times\,{\cal I}(y_{1:d}=x_{1:d})\prod_{i=1}^n{\cal N}(y_i|\,x_i,\rho^{-1})
	\prod_{i=1}^n\sum_{j=1}^\infty w_j\,{\cal N}(x_i|\,g(\t,x_{i:d}),\la_j^{-1})\prod_{i=1}^n
	{\cal N}(y_i|\,g(\t,y_{i:d}),\tau^{-1}).\nonumber
	\end{align}
Such a likelihood will lead to a Gibbs sampler with an infinite number of full conditional distributions.
To avoid that, we introduce the jointly discrete
random vectors $d^n=(d_1,\ldots,d_n)$ and $N^n=(N_1,\ldots,N_n)$ (see Ref. \cite{merkatas2017bayesian}, 
and references therein). The $d_i$ random variable, denotes the component
of the random mixture $f$ in (\ref{model1}), that the observation $x_i$ came from. In fact, 
the state space of the $d_i$ variable can be made a.s. finite, if we define
the random variable $N_i\sim f_N(\,\cdot\,|\,p)$, where $p$ is a parameter, such that,
the conditional random variable $(d_i|N_i)$ attains a discrete uniform distribution over the a.s. finite set 
${\cal S}_i=\{1,\ldots,N_i\}$. Then, it can be shown, that by letting $N_i$ to follow the particular negative 
binomial distribution $f_N(N_i|\,p)=N_i\,p(1-p)^{N_i-1}{\cal I}(N_i\ge 1)$, 
the random weights $w_j$ in (\ref{model1}), will
form the strictly decreasing geometric sequence $w_j=p(1-p)^{j-1}{\cal I}(j\ge 1)$. So that, in the
$(d^n,N^n)$-augmented posterior (\ref{posterior1}), we can switch from the variable $w^\infty$
to the variable $p$. Finally, the posterior attains the representation
	\begin{align}
	\label{posterior2}
	& \pi(p,\la^\infty,d^n,N^n,\t,x_{1:d},\tau,y^{(n)}|\,x^{(n)},\rho,{\cal R},{\cal P},{\cal M})\propto
	\pi(p,\la^\infty,\tau,\t,x_{1:d})\\
	& \quad\times\prod_{i=1\atop d_i:\,d_i\le N_i}^n p^2(1-p)^{N_i-1}
	\la_{d_i}^{1/2}\exp\left\{-{\la_{d_i}\over 2}(x_i-g(\t,x_{i:d}))^2\right\}\nonumber\\
	& \quad\times\quad {\cal I}(y_{1:d}=x_{1:d})\,
	\tau^{n/2}\exp\left\{-{1\over 2}\sum_{i=1}^n\left[
	\tau(y_i-g(\t,y_{i:d}))^2+\rho(y_i-x_i)^2\right]\right\}\nonumber.
	\end{align}
We note that, the likelihood factor in the second line of the previous equation, is very similar
to the GSBR-likelihood that appears in equation (11), of Proposition 1, 
in Ref. \cite{merkatas2017bayesian}.


\subsection{Priors and full conditional distributions} 

To complete the model, we assign independent priors to the variables $p$, $\la^\infty$, $\t$, $x_{1:d}$,
and $\tau$, namely:
\begin{enumerate}
	\item
	We set $\pi(p)={\cal B}(p|a_1,a_2)$, a beta conjugate prior, with fixed shape 
	hyperparameters $a_1$ and $a_2$.
	\item
	The variable $\la^\infty$ is an infinite sequence of independent precisions (inverse variances). 
	Nevertheless, the nonparametric MCMC will require, at each sweep, the computation of only an almost 
	surely {\it finite} number, $N^*=\max_{1\le k\le n}N_k$, of posterior $\la_j$s. 
	Standard Bayesian modeling suggests to use gamma conjugate prior distributions
	over the $\la_j$ precision parameters, so we set $\Pi(d\la^\infty)=\prod_{j=1}^\infty
	{\cal G}(\la_j|b_1,b_2)d\la_j$, where $b_1$ and $b_2$ are the fixed shape and rate hyperparameters, 
	respectively. Similarly, because $\tau$ is a precision, we set a-priori 
	$\pi(\tau)={\cal G}(\tau|\gamma_1,\gamma_2)$. 
	
	\item
	For the vector of parameters $\t=(\t_1,\ldots,\t_s)$ and for the the vector of initial conditions
	$x_{1:d}=(x_0,\ldots,x_{1-d})$, we assume the independent priors $\pi(\t)\propto 1$ and  
	$\pi(x_{1:d})\propto 1$, respectively. For example, suppose that a-priori we have
	$$
	\pi(\t_1,\ldots,\t_s)\propto\prod_{i=1}^s\exp\{-(\t_i-\t_{0,i})^2/2\sigma_{0,i}^2\},
	$$
	then letting $\sigma_{0,i}^2$ tend to infinity, one obtains $\pi(\t)\propto 1$. Such a prior 
	is noninformative, and although improper (not a density over $\R^s$), leads to a proper 
	full conditional for $\t$.
\end{enumerate}


Note that, to reduce dynamical error, the prior expectation $\E(\delta)$ will have
to be set close to zero. And if at the same time, we want to control the proximity
between the original and the noise reduced orbit, we will have to predetermine 
values for the prior means of $\ga_i$s, in the interval $[2\times 10^{-6}, 2\times 10^{-4}]$.
This is due to the fact, that the individual distances $|x_i-y_i|$ are by construction 
small. 

We have the following proposition:

\vspace{0.1in}\noindent
{\bf Proposition 1.} {\sl The full conditional distributions for the noise reduced orbit $y^n$, 
	are given by $\pi(y_j|\cdots)\propto e^{-C(y_j|\cdots)/2}$,
	where $\pi(y_j|\cdots)$ denotes the dependence of the variable $y_j$ to the rest of the variables. 
	Letting $h_\t(y_j,y_{j:d}):=(y_j-g(\t,y_{j:d}))^2$, the function $C(y_j|\cdots)$, for $j=1,\ldots,d$ 
	is given by
	\begin{align*}
	C(y_j|\cdots)&=\tau\sum_{k=0}^d h_\t(y_{j+k},y_{j+k:d})\\ 
	&\times {\cal I}(y_0=x_0,\ldots,y_{-d+j}=x_{-d+j}) +\rho(y_j-x_j)^2 ,
	\end{align*}         
	for $j=d+1,\ldots,n-d$ is given by
	$$
	C(y_j|\cdots)=\tau\sum_{k=0}^d h_\t(y_{j+k},y_{j+k:d})+\rho(y_j-x_j)^2,
	$$
	and, for $j=n-d+1,\ldots,n$, by
	$$
	C(y_j|\cdots)=\tau\sum_{k=0}^{j-n} h_\t(y_{j+k},y_{j+k:d})+\rho(y_j-x_j)^2.
	$$
}

\vspace{0.1in}\noindent{\bf Proof:} The desired result, comes from the representation
of the posterior in equation (\ref{posterior2}).\hfill $\square$


\subsection{The DNRR sampler}
To accelerate the convergence of the Gibbs
sampler based on the posterior distribution in (\ref{posterior2}), we collect our variables in to
the two groups: 
$$
G_1=\{v,\t,x_{1:d}\}\quad{\rm and}\quad G_2=\{\tau,y^n\}
$$
with $v=\{p,\la^\infty,d^n,N^n\}$.
We first sample, from the full conditionals of $G_1$ given $x^n$, and then, from the 
full conditionals of $G_2$ given $G_1$ and $x^n$. Then, it is not difficult to see that 
such a {\it blocked} Gibbs sampler scheme, admits the same stationary distribution as the plain  
Gibbs sampler scheme, coming from sampling the full conditionals of $G_1\cup G_2$ given
$x^n$, each one individually. 

\vspace{0.2in}\noindent
{\bf Proposition 2.}
{\sl  
	Given the model $\cal M$ and fixed $\rho>0$, marginally,
	$(G_2|x^n)$ is distributed as 
	\begin{equation}
	\label{scheme1}
	(G_2|x^n)\sim\int_{\R^{s+d}\times {\mathbb V}}\Pi(\,\cdot\,,\,\cdot\,|\t,x_{1:n},x^n)
	\Pi(d\t,dx_{1:n}|v,x^n)\Pi(dv|x^n),
	\end{equation}
	where $\mathbb V$ denotes the support of the random vector $v$.
}

\vspace{0.2in}\noindent{\bf Proof:} 
Given the model $\cal M$, and fixed $\rho>0$, we want to sample 
from the variable $(\tau, y^n|x^n)$. To do so, we should first sample from the joint of
$\t$ and $x_{1:n}$ given $x^n$, and then from the joint of $\tau$ and $y^n$ given 
$\t$ and $x_{1:n}$, that is
$$ 
(\t,x_{1:n}|x^n)\sim\Pi(\,\cdot\,,\,\cdot\,|x^n)
$$
and then from
$$
(\tau, y^n|x^n)\sim\Pi(\,\cdot\,,\,\cdot\,|\t,x_{1:n},x^n)
$$
whence
$$
(\tau, y^n|x^n)\sim\int_{\R^{s+d}}\Pi(\,\cdot\,,\,\cdot\,|\t,x_{1:n},x^n)\Pi(d\t,dx_{1:n}|x^n).
$$
For a generic noise source, we have to sample first from $(p,\la^\infty|x^n)$, and then 
from $(\t,x_{1:n}|p,\la^\infty,x^n)$.
However, for the creation of an a.s. finite Gibbs sampler, the random vector $(d^n,N^n)$ 
has to be introduced. Then, letting $v=(p,\la^\infty,d^n, N^n)$, one has
$$
(\t,x_{1:n}|x^n)\sim\int_{\mathbb V}\Pi(\,\cdot\,,\,\cdot\,|v,x^n)\Pi(dv|x^n),
$$
which gives the desired result.\hfill $\square$

Now, it is clear, that our model is based on the iteration of two consecutive steps, the $(\hat{g}_{x^n},\hat{f}_{x^n})$-reconstruction step and the $y^n$-sampling step: 

\begin{enumerate}		
	\item 
	We have seen that the reconstruction step, stems from the GSBR-sampler 
	introduced in Ref. \cite{merkatas2017bayesian}. The differences are: the absence of the out-of-sample 
	variables, the more general $d$-dimensional lag dependence, 
	the application of a conjugate beta prior and the application of an improper prior, on the variables 
	$p$ and $(\t,x_{1:d})$, respectively. 
	\item
	In the noise-reduction step,
	the noise reduced trajectory $y^n$, is sampled conditionally on the sampled values of the 
	reconstruction step. We can think of the reconstruction stage, as providing observations 
	from the {\it distributions} of the initial condition $y_{1:n}$ and the parameter $\t$ of the
	estimated deterministic part ${\hat g}_{x^n}$ of the new trajectory. To replicate the 
	${\hat g}_{x^n}$-dynamics, 
	under a reduced dynamical error, we use a Metropolis within Gibbs updating procedure, with a 
	small variance random walk proposal distribution, initialized at the observed $x^n$ trajectory.	
\end{enumerate}

Then, the new trajectory $y^n$, has the following properties:

\begin{enumerate}
	\item
	We define, the relative dynamical noise reduction $R_{\rm dyn}$ attained by the $y^n$ trajectory 
	with respect to the $y^n$ as
	$$
	R_{\rm dyn}(y^n,x^n;\hat{g}_{x^n}) :=  
	1 - {E_{\rm dyn}(y^n;\hat{g}_{x^n})\over E_{\rm dyn}(x^n;\hat{g}_{x^n})},
	$$
	so that $R_{\rm dyn}>r$ implies $E_{\rm dyn}(y^{(n)};\hat{g}_{x^n})<(1-r)E_{\rm dyn}(x^{(n)};\hat{g}_{x^n})$. We will see, that in all our numerical examples, it is that with $r>0.8$.
	\item
	When $\rho$ tends to infinity, the distribution of distances between the individual points 
	of the $y^n$ and $x^n$ trajectories, concentrates its mass to zero.
	\item 
	The estimated underlying deterministic parts of $y^n$ and $x^n$ are close to each other.
	For suppose, that we estimate in terms of the GSBR-sampler, the $g$-dynamics given the $x^n$ 
	and the $y^n$ trajectories. Then 
	the distance $d({\hat g}_{x^n},{\hat g}_{y^n})$ between the two deterministic parts 
	will be small; for example, when ${\hat g}_{x^n}$ and ${\hat g}_{y^n}$ are polynomials, this distance 
	could be the $l_2$-norm of the polynomial ${\hat g}_{x^n}-{\hat g}_{y^n}$. 	
\end{enumerate}


\vspace{0.2in}\noindent{\bf The sampling scheme:}
We first specify initial values for the variables $x_{1:n}$, $\t$, $\tau$, 
and we iterate for $t=1,\ldots,K$ the following sampling scheme:


\begin{enumerate}[label=S\arabic*:]
	
	\item 
	For $i=1,\ldots,n$, generate the state space range variable $N_i^{(t)}\sim\pi(N_i|\cdots)$, 
	of the allocation variable $d_i^{(t)}$.
	
	\item 
	For $i=1,\ldots,n$, generate the infinite mixture allocation variable $d_i^{(t)}\sim\pi(d_i|\cdots)$.
	
	\item
	For $i=1,\ldots,N^*$, with $N^*=\max_{1\le k\le n}N_k$, sample $\la_i^{(t)}\sim\pi(\la_i|\cdots)$.
	
	\item 
	Generate the initial condition vector $(x_{1:n})^{(t)}\sim\pi(x_{1:n}|\cdots)$ 
	
	\item 
	Generate  $\t^{(t)}\sim\pi(\t|\cdots)$.
	
	\item 
	Sample the geometric probability  $p^{(t)}\sim\pi(p|\cdots)$. 
	
	\item 
	Having updated $p^{(t)}$ and $\la^{(t)}$ up to $N^*$, sample from the noise process ${\hat f}_{x^n}$ 
	$$
	z_{n+1}^{(t)} \sim 
	\sum_{j=1}^{N^*} p^{(t)}(1-p^{(t)})^{j-1}{\cal N}\left(z_{n+1}\,|\,0,\,1/{\la_j^{(t)}}\right).		
	$$
	
	
	
	\item 
	Initialize the vector of initial conditions $(y_{1:n})^{(t)}$ of the noise reduced trajectory
	to the previously sampled initial condition $(x_{1:n})^{(t)}$ of the $x^n$, and
	iterate for $j=1,\ldots,n$ the following Metropolis-within-Gibbs sampling scheme:
	
	\begin{enumerate}
		
		\item 
		Generate proposal 
		\begin{equation}
		\label{proposal}
		y_j^*\sim y_j^{(t-1)} +\nu\,{\cal N}(0,1).
		\end{equation}
		
		\item 
		Calculate the acceptance probability $\alpha(y_j^{(t-1)},y_j^*)$ given by
		$$
		\quad\quad\min\left\{1, \exp\left\{ -{1\over2}  \left[C(y_j^*|\cdots)
		-C\left(y_j^{(t-1)}|\cdots\right)\right]\right\}\right\}.
		$$	
		
		\item 
		Accept $y_j^{(t)}=y_j^*$ with probability $\alpha(y_j^{(t-1)},y_j^*)$.
		
	\end{enumerate}	
	
	\item
	Generate $\tau^{(t)}\sim\pi(\tau|\cdots)$.
	
\end{enumerate}	

\section{Simulation Results}
In this section, we will provide numerical illustrations of the DNRR algorithm for the random
H\'enon map, and the random bistable cubic map introduced in Ref. \cite{merkatas2017bayesian}. 
In all cases, except in the case for the variable $\tau$, the prior specifications are completely
noninformative.

As a prior for the geometric probability variable, we take the arcsine density 
$p\sim{\cal B}(0.5,0.5)$, which coincides with the Jeffrey's prior for $p$. On the precisions 
$(\lambda_j)_{j\ge 1}$ of the random density $f$, we place the vague gamma prior $\lambda_j\sim{\cal G}(10^{-3},10^{-3})$, which is very close to a scale invariant prior. On the control variable $\t$, 
and the initial condition variable $x_{1:d}$, we assign the translation invariant priors $\pi(\t)\propto 1$ and $\pi(x_{1:d})\propto 1$, respectively. Because we want a-posteriori to force the
variance $\delta=\tau^{-1}$ to concentrate its mass near zero, we have to set its prior mean and variance
close to zero. We can achieve this by setting $\tau\sim{\cal G}(10^4,10^{-2})$.
Finally, to avoid mixing issues, following standard methodology, each time, we 
calibrate \cite{robert2004monte} the proposal variance 
$\nu^2$ of the embedded Metropolis-within-Gibbs sampler in equation (\ref{proposal}), such that, the 
mean acceptance probability of the sampling scheme is between 25 and 35\%.

In all our numerical experiments, the DNRR Gibbs samplers have ran for $25\times 10^4$ iterations leaving the first $5\times 10^4$ samples as a burn-in period.

\subsection{The H\'enon map} 
We consider a time series realization $x^n$ of size $n=1000$, coming from the 
random recurrence relation given in (\ref{noisyHenon})
with $e_i\IID f_{2,1}$, variance $\s^2=0.21\times 10^{-4}$ and initial condition 
$x_0=x_{-1}=0.5$ for noise level at approximately $3\%$. We model 
the deterministic part $g$, with the complete quadratic polynomial in the two variables, namely
	\begin{equation}
	\label{fqmap}
	g(\t, x_{i-1},x_{i-2})=
	\t_0+\t_1x_{i-1}+\t_2 x_{i-2}+\t_3 x_{i-1} x_{i-2}+\t_4 x_{i-1}^2+\t_5 x_{i-2}^2.
	\end{equation}


\vspace{0.2in}\noindent{\bf 1. A neutral proximity restriction:}
We first ran the DNRR sampler with the proximity parameter set to $\rho=10^2$. In fact, 
values of $\rho$ smaller than $10^4$, due to the informative nature of $\tau$, have a 
diminishing effect on the full conditional distributions of the $Y_j$ variables of proposition 1.
As a result, for such `small' $\rho$ values, the proximity restriction $\cal P$ becomes neutral, 
and the DNRR sampler estimates the noise reduced orbit $y^n$ attaining
minimum average deviation $E_{\rm dyn}$ with respect to the estimated ${\hat g}_{x^n}$, and maximum 
average distance with respect to $x^n$. 

In the first two rows of Table \ref{tfq2}, we present 
{\it percentage absolute relative errors} (PAREs) of the estimated $\t$-coefficients, with 
respect to the true values, based on the noisy and the noise reduced trajectories, of the 
maps ${\hat g}_{x^n}$ and ${\hat g}_{y^n}$, respectively. The last two columns of the 
table, display average PAREs, $\bar{\t}$, and $l^2$-distances. 
Because the $y^n$ based quantities, $\bar\t$ and $l^2$, are small, we 
consider both $x^n$ and $y^n$ based $\t$-estimations as identifying the specific H\'enon map
given in (\ref{noisyHenon}).

The posterior variance $\delta=\tau^{-1}$ has the interval 
$[1.39\times 10^{-6},1.81\times 10^{-6}]$ as a 95\% highest posterior density interval. The 
distribution of the individual variances of the $y^n$ trajectory, concentrates most of its mass in 
the interval $[0,10^{-5}]$. 

\begin{figure}
	\centering
	\includegraphics[width = 0.5\textwidth]{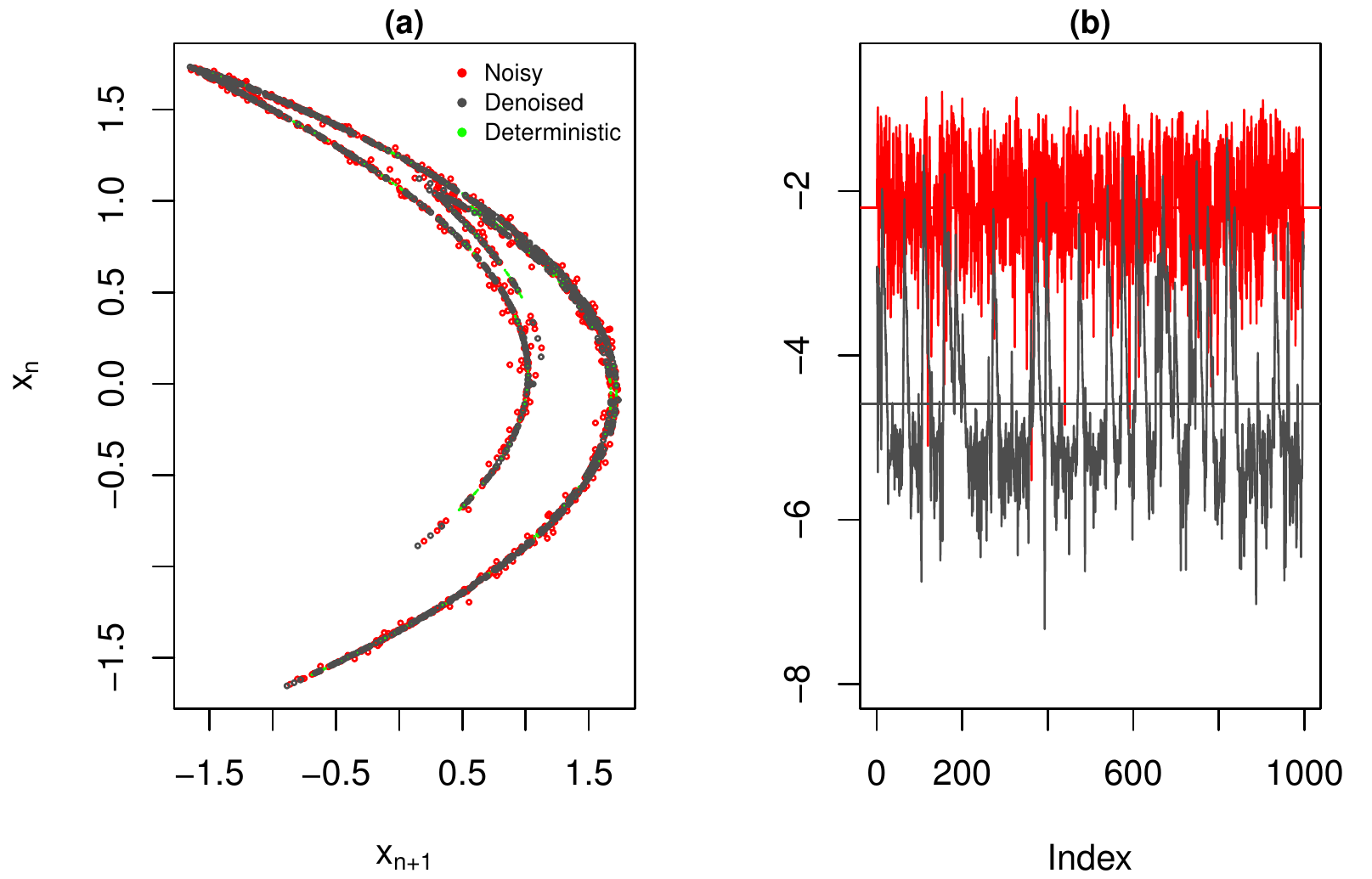}
	\caption{In figure (a), we present superimposed delay plots of the noisy, the noise reduced 
		and the deterministic trajectories of the He\'non map, of length $n=1000$.
		The associated $\log_{10}-$determinism plot is given in figure (b).}
	\label{indetplot}         
\end{figure}

\begin{figure}[H]
	\centering
	\includegraphics[width = 0.50\textwidth]{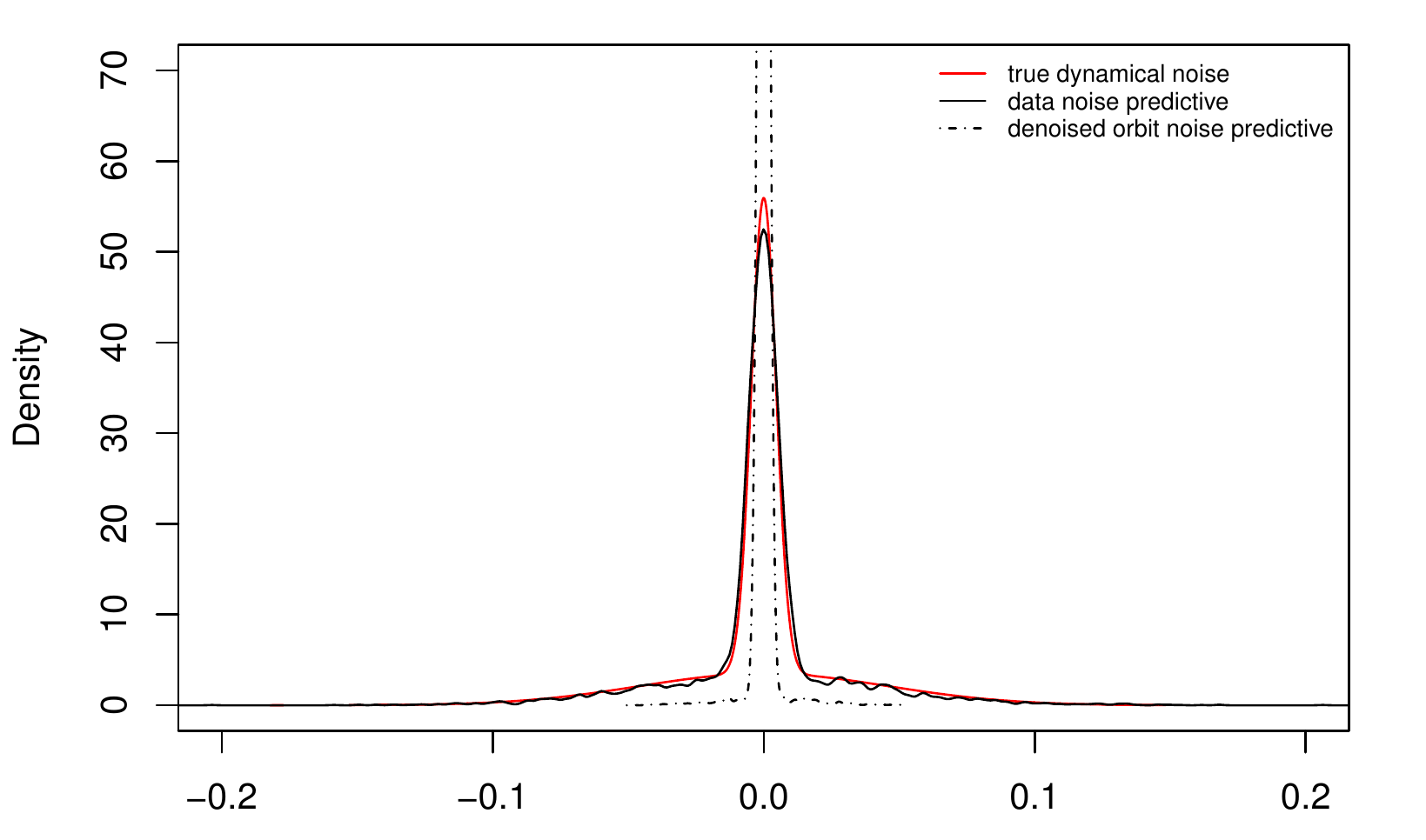}
	\caption{The true noise density $f=f_{2,1}$, for $\s^2=0.21\times 10^{-4}$, is the red continuous curve.
		Along, we superimpose the $x^n$-estimated noise density ${\hat f}_{x^n}$ as a black continuous curve,
		and the $y^n$-estimated `weaker' interactive noise density ${\hat f}_{y^n}$ as a black dashed curve.
		\label{noisepreds}}            
\end{figure}

\begin{figure*}
	\centering
	\includegraphics[width = 0.70\textwidth]{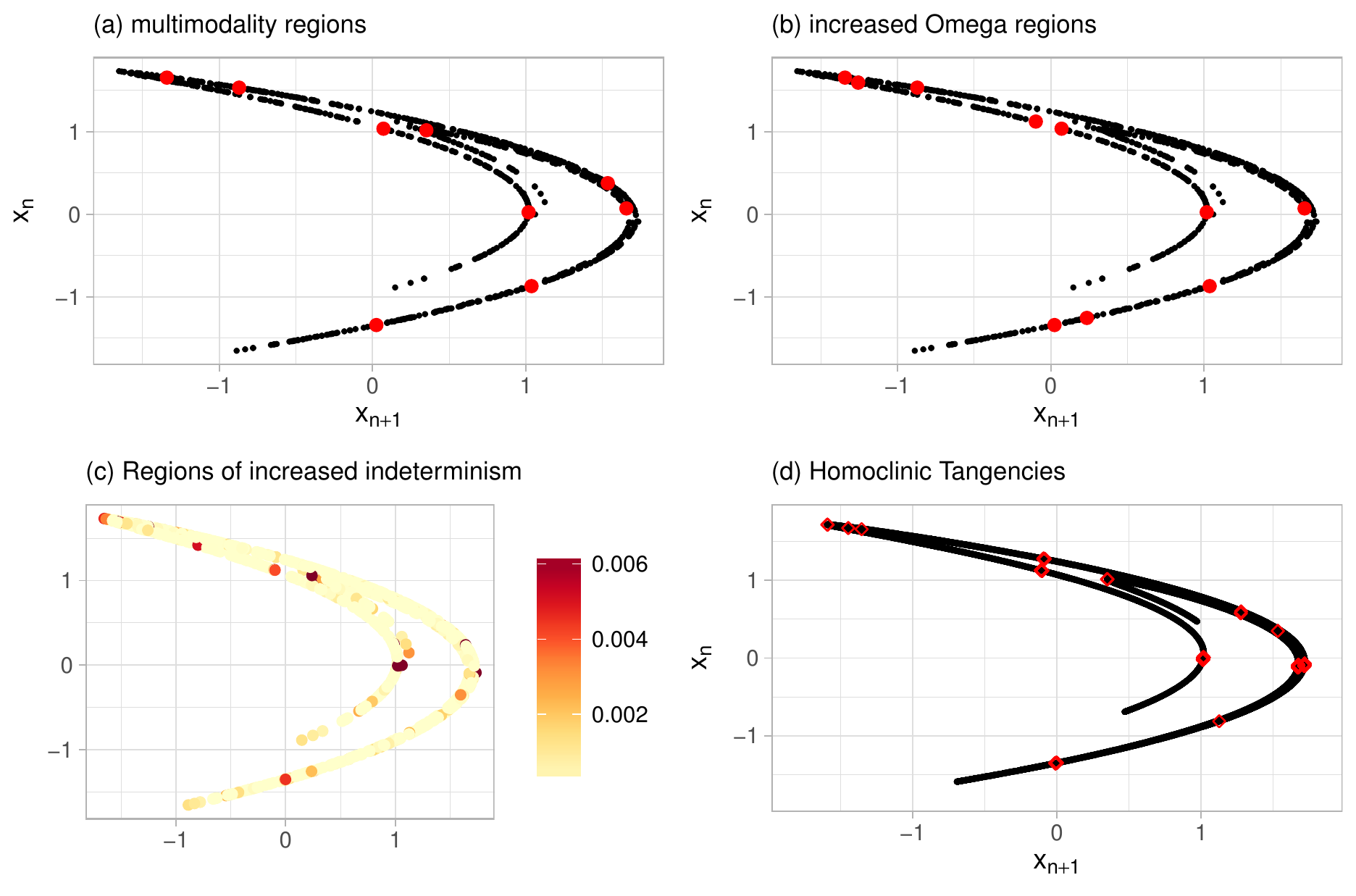}
	\caption{In Figure (a) we present a delay plot of the points in the set $M_{\rm HT}$ of the
		point estimators of the $Y_i$-posterior marginals, passing Hartigan's test for 
		unimodality. In Figure (b) we depict the delay plot of the points in the set 
		${\mathit\Omega}_{\rm HT}$ that are above the 99th percentile of the histogram of 
		${\mathit\Omega}$. Regions of high $E_{\text{dyn}}$ are depicted in Figure (c), and 
		in Figure (d) we present the primary homoclinic tangencies of the corresponding 
		deterministic attractor.} \label{hts}           
\end{figure*}

\begin{figure}
	\centering
	\includegraphics[width=0.50\textwidth]{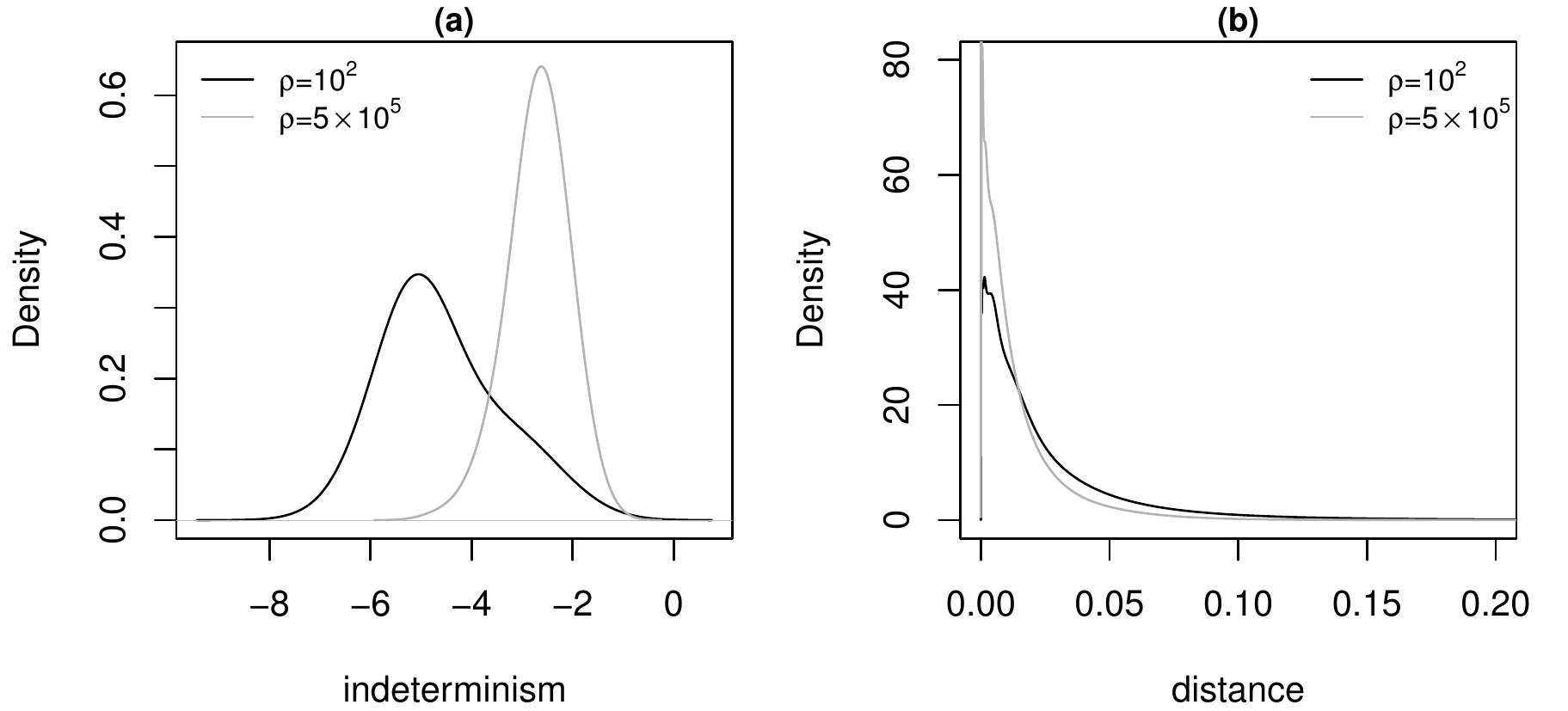}
	\caption{KDEs of (a) individual $\log_{10}$-indeterminism points and 
		(b) distance between original and noise reduced orbit points, 
		for different values of parameter $\rho$.
		\label{rhos}}            
\end{figure}

\begin{figure}
	\centering
	\includegraphics[width=0.5\textwidth]{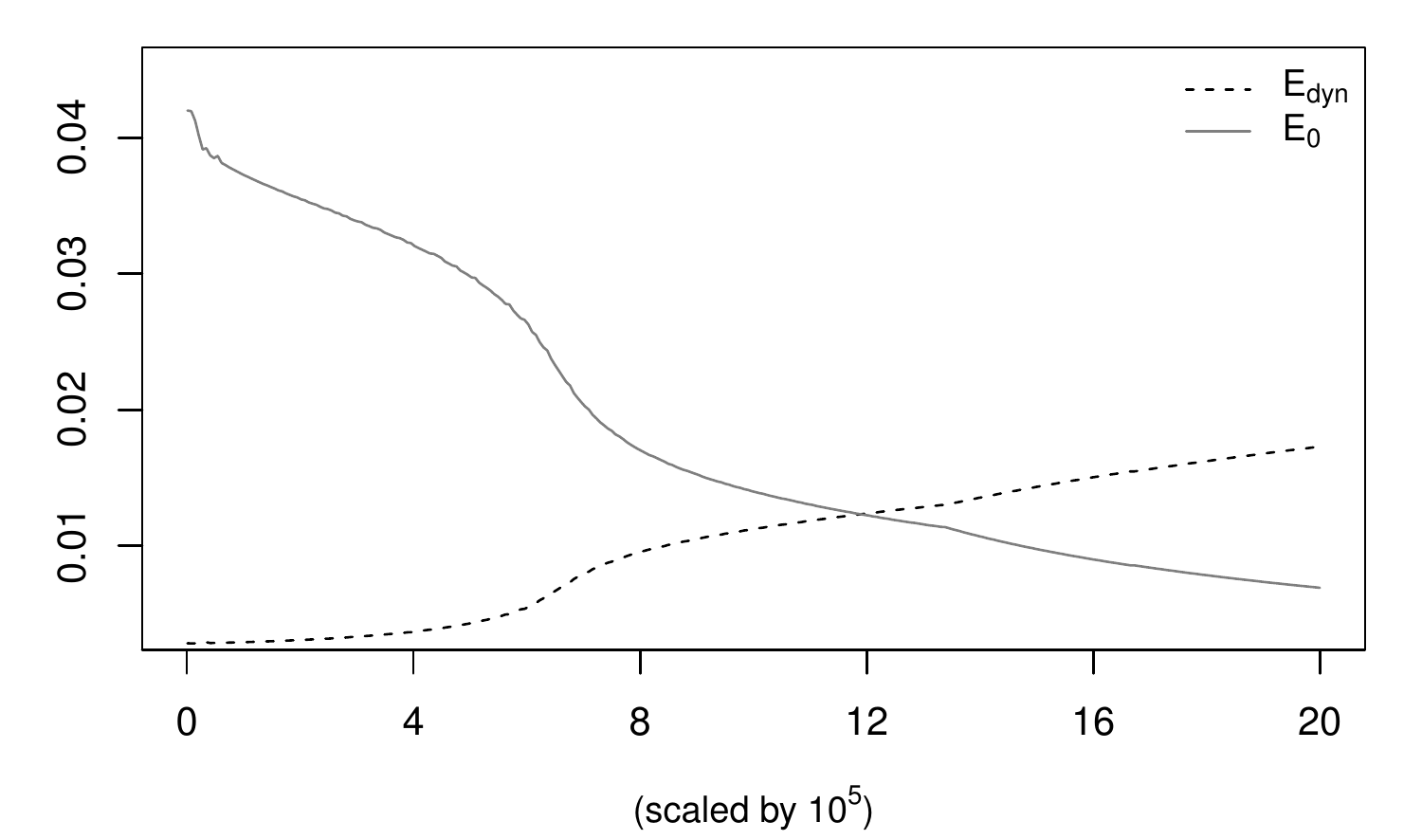}
	\caption{The average distance $E_0(y^n,x^n)$ and the average dynamic error $E_{\rm dyn}(y^n, {\hat g}_{x^n})$
		as functions of the parameter $\rho$.} 
	\label{ddr}                    
\end{figure}

We have the following results presented from Figure 
\ref{indetplot} to Figure \ref{ddr}:

\vspace{0.2in}\noindent{\bf 1.1. Noise reduction measures:}
In Figure \ref{indetplot}(a), we present superimposed the original time series $x^n$ (points in red), 
and the estimated noise reduced trajectory $y^n$ (points in dark gray) in delay coordinates.
We can see the noise reduced trajectory, shadowing the original trajectory, in the regions of 
noise-induced prolongations.
In Figure \ref{indetplot}(b), we display superimposed the individual $\log_{10}$-determinism 
plots of the original and the estimated time series, in red and dark gray color, respectively; 
for example, the individual $\log_{10}$-determinism plot of the time series $(x_i)$ is the 
trace of time series $\left(\log_{10}|E_{\rm dyn}(x_i,\hat{g})|\right)$. The red 
and black horizontal lines correspond to the average $\log_{10}$-determinisms of 
the noisy and the noise reduced times series, respectively. In the first line of Table \ref{tfq1}, 
we exhibit the denoising measures $E_{\rm dyn}$, $R_{\rm dyn}$ and $E_0$.
The average noise reduction achieved by the DNRR sampler is larger than two orders 
of magnitude, with $R_{\rm dyn}(y^n,x^n;\hat{g}_{x^n}) = 0.902$, 
$E_{\rm dyn}(y^n;\hat{g}_{x^n}) = 0.00286$ and $E_0(x^n,y^n) = 0.0428$.

\begin{table*}
	\centering
	\caption{Relative dynamical noise reductions, average indeterminisms and average distances,
		for two different values of $\rho$.}
	\label{tfq1}		    
	\begin{tabular}{ccccc}  
		$\rho$        & $E_{\rm dyn}(x^n,\hat{g}_{x^n})$&$E_{\rm dyn}(y^n,\hat{g}_{x^n})$  
		& $R_{\rm dyn}$ &    $E_0$                                \\
		\hline
		$10^2$        & 0.02932 &     0.00286    &     0.9023     &	  0.0428	\\
		$5\times 10^5$& 0.02932 &     0.00710    &     0.7577	   &	  0.0223	\\                  
		\hline
	\end{tabular}
\end{table*}

\begin{table*}
	\centering
	\caption{PAREs, average PAREs and $l^2$-distances, for the estimated coefficients of the 
		deterministic part of the perturbed H\'enon map in (\ref{noisyHenon}), based on 
		the noisy and the corresponding noise reduced trajectories, for two different values 
		of $\rho$.}
	\label{tfq2}	    	
	\begin{tabular}{cccccccccc}  
		Time series& $\rho$ & $\t_0$ & $\t_1$& $\t_2$ & $\t_3$ & $\t_4$ & $\t_5$ & $\bar\t$ & $l^2$\\
		\hline
		$x^n$	& $10^2$ & 0.089& 0.096 &	0.046	& 0.044	& 0.011 & 0.070	& 0.059	& 0.00177\\
		$y^n$	&        & 0.063& 0.043	&	0.022 & 0.028	& 0.020	& 0.038	&0.036	& 0.00110\\
		\hline
		$x^n$ & $5\times 10^5$& 
		0.079& 0.071	&	0.041	& 0.031	& 0.002	& 0.059	&0.047	& 0.00146\\
		$y^n$	&               &  
		0.177& 0.155	&	0.015	& 0.023	& 0.005	& 0.157	&0.089	& 0.00330\\               
		\hline
	\end{tabular}
\end{table*}


\vspace{0.2in}\noindent{\bf 1.2. Dynamic noise estimation:}
In Figure \ref{noisepreds}, we display superimposed the true noise density $f=f_{2,1}$ (red continuous curve), 
the $x^n$ based estimated noise density ${\hat f}_{x^n}$ (black continuous curve) and the $y^n$
based estimated noise density ${\hat f}_{y^n}$ (black dashed curve). We remark the closeness of the
noise densities $f$ and ${\hat f}_{x^n}$, and the fact that 
the ${\hat f}_{y^n}$ density, represents a much `weaker' error process. The latter, along with 
the fact that the $\t$-estimation based on 
the noise reduced trajectory identifies the specific H\'enon map, validates our contention, that the 
noise reduced trajectory comes from a dynamical system very close to the original one, 
perturbed interactively by a `weaker' error process.

\vspace{0.2in}\noindent{\bf 1.3. The existence of HTs as a cause for a-posteriori multimodality:}
While most of the $Y_i$-MPDs are unimodal, 
a small number of them is multimodal, namely, those that their support contains the projection
of a point of HT.
We have used the Hartigan's statistical test \cite{hartigan1985dip} for multimodality, 
to choose the appropriate $Y_i$-point estimator; we utilize the 
maximum a-posteriori (MAP) estimator for the case of a $Y_i$-multimodal MPD, 
and the sample mean estimator for the unimodal case.
In Figure \ref{hts}(a) we present a delay plot of the set $M_{\rm HT}$ of MAP estimations 
(solid red circles) coming from the $Y_i$-posterior marginals, passing the Hartigan's test 
for multimodality. Alternatively, we could consider the $Y_i$-predictive-samples, coming from the 
embedded Metropolis-within-Gibbs sampler, after burn-in. For each $Y_i$-sample, 
we compute the forecastable component analysis index ${\mathit\Omega}_i$ \cite{Goerg, GoergForeCA}, 
which is normalized in the interval $[0,1]$.
We let ${\mathit\Omega}=\{{\mathit\Omega}_i:1\le i\le n\}$, and we consider the subset of points 
${\mathit\Omega}_{\rm HT}$ of ${\mathit\Omega}$, that are above the 99th
percentile of its histogram, and thus, their predictive distribution exhibits {\it more structure}. 
In Figure \ref{hts}(b) we depict a delay plot of ${\mathit\Omega}_{\rm HT}$ (solid red circles).
We can see that the points in the sets $M_{\rm HT}$ and ${\mathit\Omega}_{\rm HT}$
are related to the areas of increased indeterminism depicted in Figure \ref{hts}(c). The location 
of the deterministic primary HTs are given in Figure \ref{hts}(d). 
We remark that the sets $M_{\rm HT}$ and ${\mathit\Omega}_{\rm HT}$, for fixed $n$, 
are random (point process realizations)
because they depend on the particular realization of the time series $x^n$, for example 
$\omega\mapsto{\mathit\Omega}_{\rm HT}={\mathit\Omega}_{\rm HT}(y^n|x^n(\omega))$.

\vspace{0.2in}\noindent{\bf 2. The average distance $E_0$ as a function of $\rho$:}
Here we perform a series of executions of the DNRR sampler with the same prior set up, 
and the same observed time series $x^n$, as in the previous subsection, for different values of
the $\rho$ parameter. We have taken $\rho\in\{\rho_j=j\times 10^4:j=1,\ldots,200\}$. For 
example, for $\rho=5\times 10^5$, the effect of the proximity 
restriction becomes very strong. In the second line of Table \ref{tfq1}, we present the 
noise reduction measures $E_{\rm dyn}$, $R_{\rm dyn}$ and $E_0$. The average noise reduction 
achieved in this case decreases to$R_{\rm dyn}(y^n,x^n;\hat{g}_{x^n}) = 0.7577$. The average 
indeterminism of $y^n$ with respect to $\hat{g}_{x^n}$ escalates to 
$E_{\rm dyn}(y^n;\hat{g}_{x^n}) = 0.00710$, 
with the average distance decreased considerably to $E_0(x^n,y^n) = 0.0223$.
In Figure \ref{rhos}(a), we present superimposed, the distributions of the individual 
$\log_{10}$-indeterminisms of the noise reduced trajectory with respect to $\hat{g}_{x^n}$, 
for $\rho=10^2$ (curve in black) and $\rho=5\times 10^5$ (curve in grey). 
We can see that for large values of $\rho$ the density of $\log_{10}$-indeterminisms becomes
more peaked and shifts to the right. In Figure \ref{rhos}(b), the density of the individual 
distances for the large value of $\rho$ concentrates its mass near zero.

In Figure \ref{ddr}, we present the noise reduction measures $E_{\rm dyn}(y^n,\hat{g}_{x^n})$ and
$E_0(y^n,x^n)$ as functions of $\rho$. It is that as $\rho$ increases, the average
indeterminism and the average distance are increasing and decreasing, respectively.

\vspace{0.2in}\noindent{\bf 3. Fixed noise levels imply fixed relative noise reduction:}						
In this experiment we choose the variances and the time series realizations 
$x^n$, for each $f_{2,l}$ noise process for $1\le l\le 4$, such that, they give an 
associated noise level $\eta$ of about 3\%. In the fourth column of Table \ref{tfq3}, we can see 
that the relative noise reduction measure $R_{\rm dyn}$, does not undergo major changes, 
and it attains values between 0.871 and 0.902.

\begin{table}
	\centering
	\caption{Measures of reconstruction and noise reduction efficiency for the $f_{2,l}$ 
		noise processes. The variances of the noise processes, and each realization 
		has been chosen, such that, $\eta$ is fixed at about 3\%, where
		$E_{\rm dyn}=E_{\rm dyn}(y^n,\hat{g}_{x^n})$.}
	\label{tfq3}	     	
	\begin{tabular}{ccccccc}            
		Noise    & $\sigma^2\times 10^{4}$ &   $E_0$   &  $E_{\rm dyn}$  & $R_{\rm dyn}$
		& $\bar{\t}_{x^n}$ & $\bar{\t}_{y^n}$                     \\
		\hline
		$f_{2,1}$  & $0.21$ & 0.0428   &  0.00286  & 0.902 & 0.059  & 0.036  \\               
		$f_{2,2}$  & $0.29$ & 0.0514   &  0.00371  & 0.871 & 0.115  & 0.062  \\    
		$f_{2,3}$  & $0.40$ & 0.0490   &  0.00392  & 0.871 & 0.072  & 0.098  \\               
		$f_{2,4}$  & $0.77$ & 0.0627   &  0.00323  & 0.892 & 0.054  & 0.059  \\  
		\hline		
	\end{tabular}
\end{table} 


\subsection{A bistable cubic map} 
Here, we consider the cubic map
\begin{equation}
\label{bimap}
x_i=g(\vartheta,x_{i-1})=0.05+\vartheta x_{i-1}-0.99 x_{i-1}^3.
\end{equation}
For $\vartheta\in\Theta_{\rm bi}=[1.27, 2.54]$ the map is bistable in the sense that
two mutually exclusive period-doubling cascades coexist. For values of $\vartheta$
close to 2.54, we denote the two coexisting attractors by ${\cal O}_1\subset I_1$ 
and ${\cal O}_2\subset I_2$, with approximately $I_1=[-1.60,-0.10)$ and $I_2=[-0.10, 1.67]$.
For values of $\vartheta$ slightly larger than 2.54, the set ${\cal O}_2$
undergoes a sudden change. It becomes repelling, and all trajectories over $I_1\cup I_2$ 
are attracted by ${\cal O}_1$. In fact, similar behavior can be observed for all 
$\vartheta\in(2.54,2.65)$. 

We let $\vartheta=\vartheta^*=2.55$ and we consider the 
dynamically perturbed map $x_i=g(\vartheta^*,x_{i-1})+e_i$ with $e_i\IID f_{2,1}$,
$\sigma^2=0.55\times 10^{-4}$, and $\rho=10^2$.
Then, {\it noise-induced jumps} are taking place between the intervals $I_1$ and $I_2$. 
Here we consider dynamically perturbed time series observations $x^n$, of small sample size $n=200$.
As a modeling polynomial,  we utilize the general quintic polynomial 
$g(\t, x_{i-1})=\sum_{k=0}^5\t_j x_{i-1}^k$.

\begin{figure}[H]
	\centering
	\includegraphics[width = 0.5\textwidth]{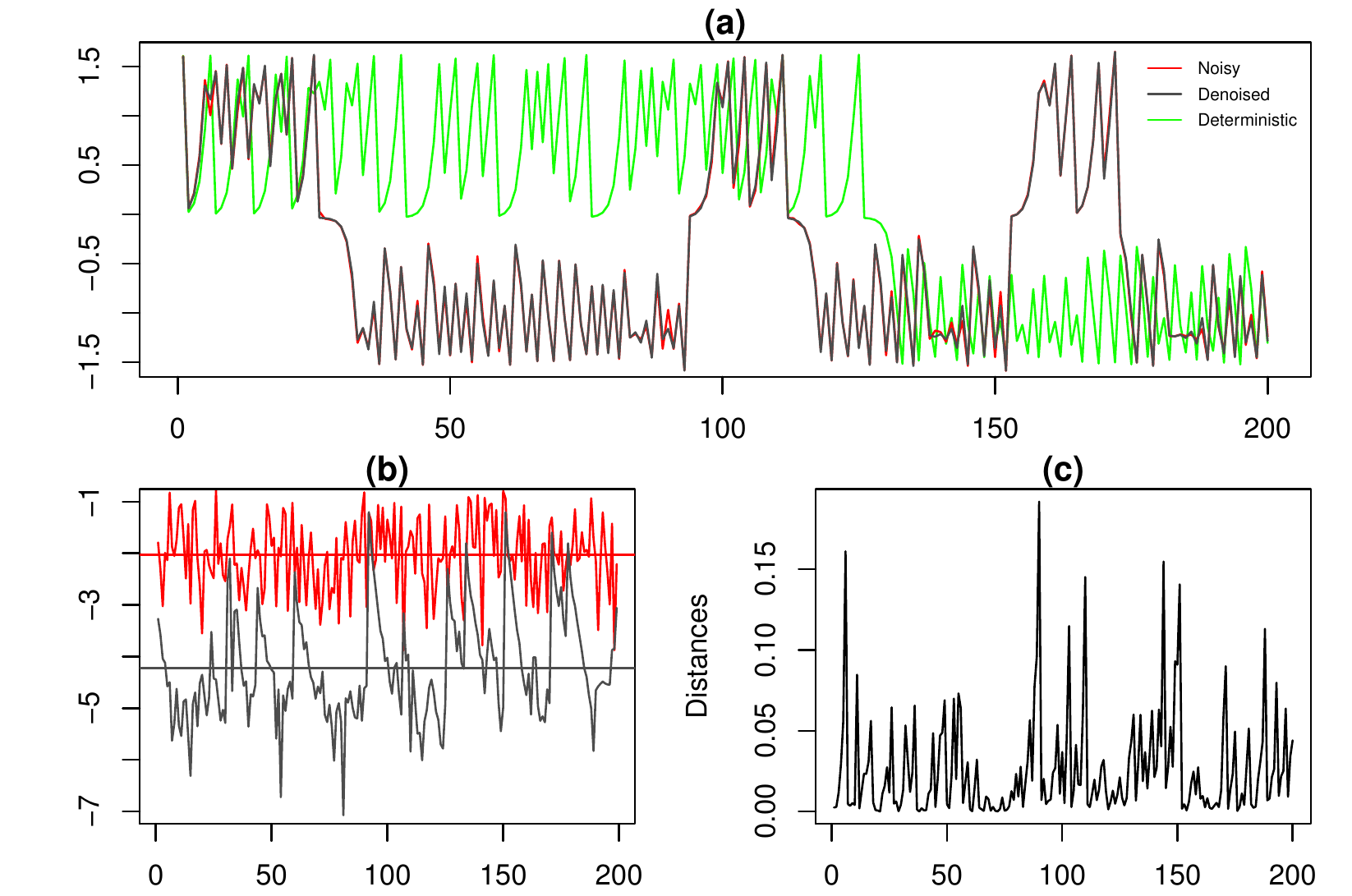}
	\caption{In Figure (a), we give superimposed, the deterministic trajectory,
		the noisy trajectory $x^n$ and the estimated $y^n$ trajectory.
		In Figure (b) we present the corresponding $\log_{10}$ indeterminism plot. The trace of
		the individual distances between the original and the noise reduced trajectory is given in 
		Figure (c).\label{polyfig}}            
\end{figure}

\begin{figure}[H]
	\centering
	\includegraphics[width=0.50\textwidth]{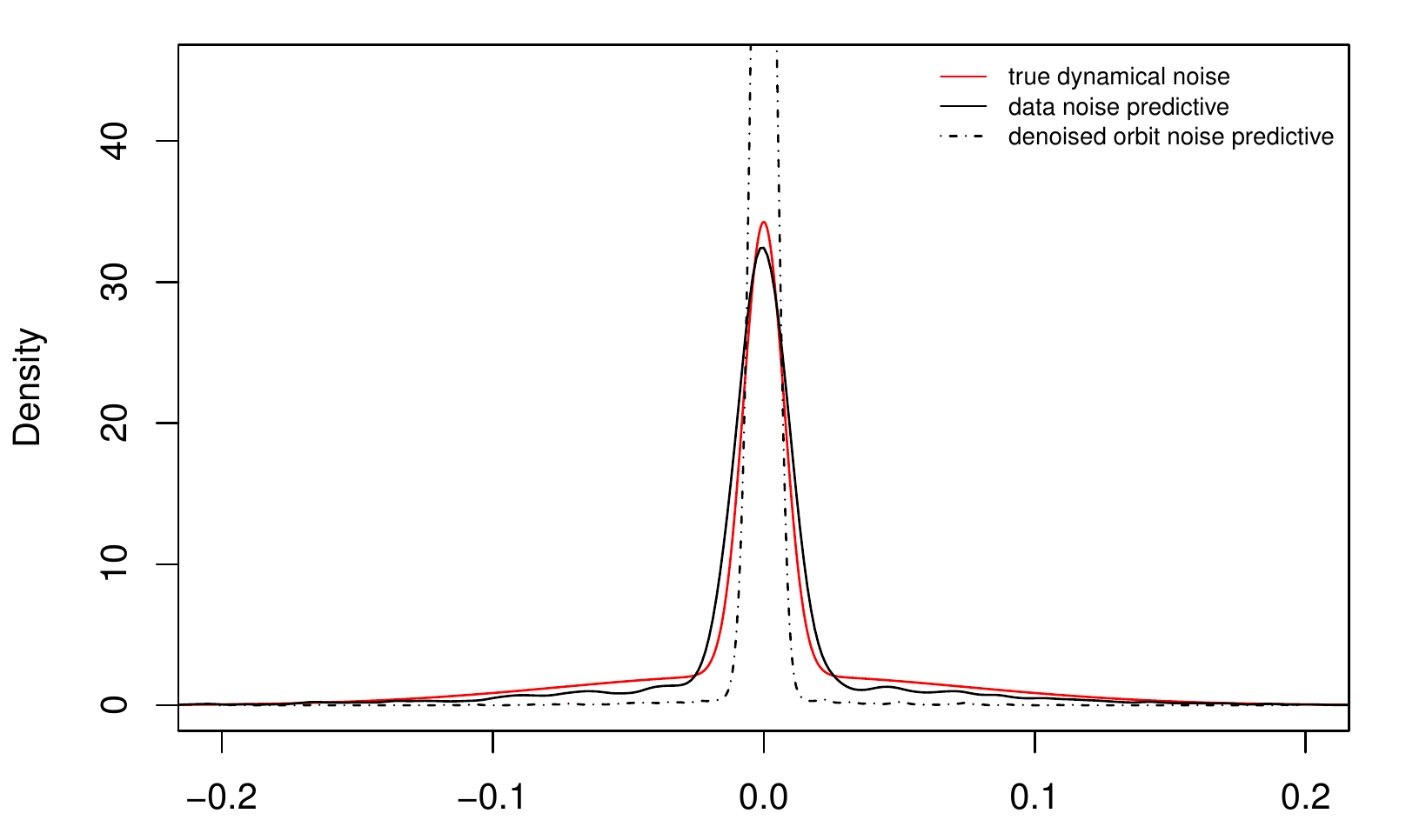}
	\caption{Kernel density estimations based on the predictive samples of ${\hat f}_{x^n}$ 
		(continuous black curve), the predictive samples of ${\hat f}_{y^n}$ (dashed 
		black curve) along with the true dynamical noise density (continuous red curve).\label{polynoise}}            
\end{figure}

\vspace{0.2in}\noindent{\bf Noise reduction in the neighborhood of noise induced jumps:}
In Figure \ref{polyfig}(a), we can see the estimated $y^n$ trajectory (in black)
evolving in the neighborhood of the original trajectory $x^n$ (in red), incorporating 
the weaker dynamical noise ${\hat f}_{y^n}$, given in Figure \ref{polynoise}, as a 
black dashed density. We remark, that our method, is based on the fact 
that it allows only small stochastic steps around the original orbit, and thus, the noise 
reduced orbit follows closely the original orbit even to its noise-induced prolongations 
in the interval $I_2$. The corresponding $\log_{10}$ indeterminism plot is given in Figure 
\ref{polyfig}(b). The plot of the individual distances between the original and the noise 
reduced trajectory is given in Figure \ref{polyfig}(c). 
In Table \ref{polytable} we display the noise reduction efficiency for the cubic
map, for noise levels between 3.5\% and 7.5\%. In the last column of the table are displayed the average 
PAREs ${\bar\theta}_{y^n}$ of the $y^n$ based estimation of the deterministic part of the noise reduced 
dynamics. We have observed, that the average PARE becomes larger than 1\%, when the noise level exceeds 8\%. 

\begin{table}
	\centering
	\caption{Measures of reconstruction and noise reduction efficiency for the cubic map,
		for various $\sigma^2$'s for the $f_{2,1}$ noise processes, where
		$E_{\rm dyn}=E_{\rm dyn}(y^n,\hat{g}_{x^n})$.}
	\label{polytable}	
	\begin{tabular}{ccccccc}
		$\sigma^2\times 10^{4}$ & \,\,$\eta$\,\,\% &  $E_0$  &$E_{\rm dyn}$ & $R_{\rm dyn}$
		& ${\bar\t}_{x^n}$ & ${\bar\t}_{y^n}$\\
		\hline
		$0.33$ & 3.5  & 0.0395   &  0.00749  & 0.812 & 0.281 & 0.425\\  			
		$0.55$ & 4.5  & 0.0413   &  0.00695  & 0.842 & 0.605 & 0.804\\
		$0.59$ & 5.5  & 0.0631   &  0.00952  & 0.826 & 0.438 & 0.262\\      
		$0.67$ & 6.5  & 0.0453   &  0.00847  & 0.848 & 0.872 & 0.958\\    
		$1.00$ & 7.5  & 0.0630   &  0.00819  & 0.867 & 0.856 & 0.987\\   
		\hline
	\end{tabular}
\end{table}

\section{Discussion} 
We have presented, a novel approach to the problem of noise reduction of dynamically 
perturbed nonlinear maps, the DNRR sampler. Our approach is Bayesian, modeling a noise 
reduced trajectory $y^n$,
that {\it evolves in the neighborhood} of a given noisy trajectory $x^n$. Our proposed DNRR algorithm,
is flexible and accurate, because the assumptions for the underlying noise process $f$ 
perturbing the original trajectory are relaxed.  
A-priori, we consider the noise as coming from a random countable mixture of zero mean Gaussians.
Then, the number of the components, the weights, and the variances of the normal mixture ${\hat f}_{x^n}$,
approximating the actual noise process $f$, are estimated directly from the observed time series.
This in turn, implies a high accuracy estimation 
of the deterministic part ${\hat g}_{x^n}$, which is the basic ingredient of the replication part 
of the DNRR sampler. Also, we have seen, that for moderate noise levels, 
the noise reduced trajectory $y^n$, has an estimated deterministic part ${\hat g}_{y^n}$ remaining
close to the estimated deterministic part ${\hat g}_{x^n}$ of the original trajectory.

We could modify the proposed DNRR model, by dropping the assumption of a known functional form for the 
deterministic part, and instead, apply over $g$, a Gaussian Process prior \cite{rasmussen2004gaussian}. 
We believe, that such an approach, will be appropriate for a wide variety of real world data sets, characterized by strong nonlinearity and (or) complicated contaminating dynamic noise.

\bibliographystyle{plain}
\bibliography{denoising}

\end{document}